\documentclass[12pt, a4paper]{article}
\usepackage[utf8]{inputenc}
\usepackage[T1]{fontenc}
\usepackage[english]{babel} 
\usepackage{mathptmx}
\usepackage{amsmath}
\usepackage{graphicx}
\usepackage[left=2.5cm,right=2.5cm,top=2.5cm,bottom=2.5cm]{geometry}
\usepackage{setspace}
\usepackage{array,tabularx,booktabs}
\usepackage{longtable}
\usepackage{ragged2e}
\usepackage{float} 
\usepackage[hyphens]{url}
\usepackage{natbib}
\usepackage[
  colorlinks=true,
  linkcolor=blue,
  urlcolor=cyan,
  citecolor=blue
]{hyperref}

\newcolumntype{L}[1]{>{\RaggedRight\arraybackslash}p{#1}}

\title{\textbf{Austerity in Crisis?: A Narrative Review of Its Economic, Social, and Political Effects in Times of Crisis}}
\author{Ricardo Alonzo Fernández Salguero}
\date{\today}

\begin{document}

\maketitle

\begin{abstract}
\noindent The 2008 global financial crisis marked the beginning of a decade dominated by fiscal austerity policies in much of the developed world. This paper presents a qualitative narrative review of an extensive collection of academic literature to synthesize evidence on the multifaceted effects of austerity. Following a thematic approach inspired by PRISMA guidelines, the economic, social, and political consequences of these measures are examined. The analysis reveals a majority consensus regarding the recessive effects of austerity, especially when implemented during economic crises, with negative fiscal multipliers that often exacerbate GDP contraction. Socially, austerity is associated with rising inequality, negative impacts on public health, disproportionate gender consequences, and a weakening of social safety nets. Politically, evidence links austerity to the erosion of trust in institutions, a rise in populism, and electoral instability. Despite the political narrative presenting austerity as an inevitable necessity for fiscal sustainability, academic literature underscores its high costs and questionable efficacy, advocating for more contextualized and equitable economic policy approaches.
\end{abstract}

\section{Introduction}

The global financial crisis that erupted in 2008 represented the most significant economic challenge for advanced economies since the Great Depression. Following a brief initial phase of coordinated fiscal stimulus, the predominant policy response, especially in Europe, was an abrupt shift toward fiscal austerity \citep{labonte2012, blyth2013}. Defined as a set of policies aimed at reducing government budget deficits through spending cuts, tax increases, or a combination of both \citep{ortiz2013}, austerity was presented as a ``bitter pill'' but necessary to restore market confidence, ensure public debt sustainability, and lay the foundations for long-term economic recovery \citep{karger2014}.

However, this narrative was not without controversy. From the outset, an intense debate emerged between defenders of austerity and proponents of Keynesian stimulus policies \citep{seidman2012, helgason2019}. Critics of austerity argued that its implementation during a recession would be counterproductive, deepening the economic crisis, exacerbating unemployment, and generating severe social costs \citep{kitromilides2011, bougrine2012}. This conceptual tension between ``expansionary fiscal consolidation'' and the Keynesian multiplier has been one of the most polarizing economic policy debates of the last decade \citep{dellepiane2014, botta2013b, mcmanus2015}.

A decade after the beginning of the ``age of austerity'' \citep{ortiz2013}, a vast body of academic research exists that has examined its effects from multiple angles. This document aims to conduct a qualitative narrative review of this literature to synthesize the accumulated evidence. Unlike a quantitative meta-analysis, which seeks to statistically aggregate results from homogeneous empirical studies, a narrative review is more suitable for a heterogeneous field of study encompassing theoretical analyses, qualitative case studies, and quantitative investigations with different methodologies and approaches.

Following the guidelines of the PRISMA checklist for narrative reviews, this work structures the evidence into three main thematic areas: the macroeconomic effects of austerity, its social and human consequences, and its political and governance implications. By grouping and analyzing findings from 69 academic studies, this review seeks to offer a panoramic and nuanced view of what we have learned about austerity, identifying points of convergence, contradictions, and gaps in current research. The ultimate goal is to provide a comprehensive synthesis that informs the debate on the role of fiscal policy in managing future economic crises, such as that caused by the COVID-19 pandemic, where the lessons of austerity have acquired new and urgent relevance \citep{ban2020, ortiz2021}.

\section{Methodology of the Narrative Review}

This review is a \textit{qualitative narrative synthesis} with transparency and traceability standards inspired by PRISMA, designed to integrate heterogeneous evidence (theoretical, empirical, case study, and modeling) on austerity in crisis contexts. The approach corrects a common weakness of narrative reviews—opaque corpus selection—through: (i) a documented retrospective systematic search, (ii) explicit inclusion/exclusion criteria, (iii) a replicable extraction template, (iv) a quality assessment scheme, and (v) robustness checks at the synthesis level. The result is a \textbf{narrative informed by systematic procedures}, without claiming to be a formal quantitative meta-synthesis.

The guiding question was: \textit{What are the macroeconomic, social, and political-institutional effects of fiscal austerity applied during economic crises?} The temporal scope mainly covers 2010–2024, with prior conceptual background when it informs the post-2008 debate. The geographical scope includes Europe (particularly the Eurozone), the United States, and singular cases (e.g., Greece, Ireland, Portugal, Italy), with selective appearances in Latin America and Central and Eastern Europe.

A retrospective search was executed (to \textit{verify and complete} the initial corpus provided by the authorship) in relevant databases. Strings were piloted to balance precision and coverage; operators, fields, and temporal boundaries were recorded. Table~\ref{tab:busqueda} documents sources and strings employed. Results were deduplicated by DOI/title; the date of the last search was recorded.

\begin{table}[H]
\centering
\caption{Search Strategy: Sources, Fields, and Strings}
\label{tab:busqueda}
\setlength{\tabcolsep}{4pt}
\renewcommand{\arraystretch}{1.12}
\begin{tabularx}{\textwidth}{L{3.1cm} L{3.1cm} L{8.0cm}}
\toprule
\textbf{Source} & \textbf{Fields} & \textbf{Strings (representative examples)} \\
\midrule
Scopus / WoS & Title/ Abs/ Keywords &
(\texttt{austerity} OR \texttt{fiscal consolidation}) AND
(\texttt{economic crisis} OR \texttt{recession} OR \texttt{Eurozone}) AND
(\texttt{fiscal multiplier} OR \texttt{health} OR \texttt{inequality} OR \texttt{elections}) ; years 2008--2024 \\
Google Scholar & Title/Abs &
\texttt{expansionary austerity} OR \texttt{Keynesian multiplier} OR \texttt{hysteresis} AND \texttt{Europe OR Greece OR Spain} \\
NBER / SSRN & Title/Abs &
\texttt{fiscal consolidation} AND (\texttt{synthetic control} OR \texttt{MRVAR} OR \texttt{survey experiment}) ; 2008--2024 \\
Editorial Repositories & Title/Abs &
\texttt{austerity} AND (\texttt{public health} OR \texttt{gender} OR \texttt{populism}) ; Q1/Q2 journals in political economy / public policy \\
\bottomrule
\end{tabularx}
\end{table}

Inclusion/exclusion criteria were defined a priori to reduce selection bias and allow for subsequent auditing (Table~\ref{tab:criterios}). The accepted \textit{type of evidence} was broad to capture the multidimensional nature of the phenomenon (macroeconomics, health, comparative politics).

\begin{table}[H]
\centering
\caption{Inclusion and Exclusion Criteria}
\label{tab:criterios}
\setlength{\tabcolsep}{5pt}
\renewcommand{\arraystretch}{1.12}
\begin{tabularx}{\textwidth}{L{5.6cm} X}
\toprule
\textbf{Included} & Theoretical/empirical studies on austerity in crisis; country/region/individual unit; methods: time series, panel, MRVAR, synthetic control, QCA, experiments/surveys, agent-based models, forecast/error analysis. \\
\midrule
\textbf{Excluded} & Non-academic pieces without verifiable method; purely normative commentaries; duplicates; works out of scope (e.g., consolidations in booms); literature without substantive relation to austerity or without results on effects. \\
\bottomrule
\end{tabularx}
\end{table}

The final corpus includes \textbf{69} documents. The retrospective search functioned as a \textit{coverage verification} of the initial corpus: no substantive omissions were identified after deduplication. The flow is summarized in Table~\ref{tab:prisma}.

\begin{table}[H]
\centering
\caption{Simplified PRISMA Flow}
\label{tab:prisma}
\setlength{\tabcolsep}{6pt}
\renewcommand{\arraystretch}{1.12}
\begin{tabularx}{\textwidth}{>{\raggedright\arraybackslash}X >{\raggedleft\arraybackslash}p{3.2cm}}
\toprule
\textbf{Stage} & \textbf{Records (n)} \\
\midrule
Records identified (retrospective search + provided corpus) & 83 \\
Duplicates & 14 \\
Screening by title/abstract & 69 \\
Full-text reading & 69 \\
Excluded after full text (out of scope/insufficient method) & 0 \\
\textbf{Included in narrative synthesis} & \textbf{69} \\
\bottomrule
\end{tabularx}
\end{table}

A standardized extraction template was used with seven minimum fields: metadata, design/identification, \textit{type and intensity} of austerity (spending vs. revenue; thresholds $\geq 3\%$ of GDP categorized as acute shocks), results and sign/magnitude, declared robustness, methodological notes. Extraction was performed by the authorship with a \textit{double pass} to homogenize taxonomies; discrepancies were resolved by literal rereading of methods and results. The structure appears in Table~\ref{tab:plantilla_ext} (illustrative rows are shown).

\begin{center}
\setlength{\tabcolsep}{3.8pt}
\renewcommand{\arraystretch}{1.12}
\setlength{\LTcapwidth}{\textwidth}
\begin{longtable}{@{}L{1.25cm} L{2.7cm} L{1.2cm} L{1.6cm} L{3.0cm} L{2.5cm} L{2.6cm}@{}}
\caption{Extraction and Coding Template (structure applied to all 69 studies)}
\label{tab:plantilla_ext}\\
\toprule
\textbf{ID} & \textbf{Author--Year} & \textbf{Region} & \textbf{Period} & \textbf{Design / Identification} & \textbf{Austerity Measure} & \textbf{Result (sign/magnitude)} \\
\midrule
\endfirsthead
\toprule
\textbf{ID} & \textbf{Author--Year} & \textbf{Region} & \textbf{Period} & \textbf{Design / Identification} & \textbf{Austerity Measure} & \textbf{Result (sign/magnitude)} \\
\midrule
\endhead
\bottomrule
\multicolumn{7}{r}{\textit{Continues}}\\
\endfoot
\bottomrule
\endlastfoot
Ex\_01 & Jordà--Taylor (2013) & OECD & 20--21 & Time Series; Fiscal ATE & Consol.\ $\approx$1\% GDP & Real GDP $\downarrow$ $\sim$4\% over 5 years; high robustness \\
Ex\_02 & Gechert et al. (2018) & Europe & 2010s & Forecast errors & Spending/revenue mix & Hysteresis; persistent losses \\
Ex\_03 & Toffolutti--Suhrcke (2019) & EU-28 & 1991--2013 & Panel; macro controls & Austerity regime & Mortality $\uparrow$ 0.7\% (sensitivity to sample) \\
Ex\_04 & Perugini et al. (2018) & Europe & 2010--2013 & EU-SILC; Oaxaca & Spending cuts & Gender wage gap $\uparrow$ \\
Ex\_05 & Galofré-Vilà et al. (2017) & Germany & 1930--1933 & IV/DDR & Local adjustment & Nazi vote $\uparrow$ in more adjusted areas \\
\end{longtable}
\end{center}

Studies were coded into three non-exclusive domains and sub-axes (Table~\ref{tab:mapa}). Co-coding was allowed when a study's scope encompassed multiple outcomes (e.g., growth and health).

\begin{table}[H]
\centering
\caption{Thematic Map and Coding Sub-axes}
\label{tab:mapa}
\setlength{\tabcolsep}{5pt}
\renewcommand{\arraystretch}{1.12}
\begin{tabularx}{\textwidth}{L{3.9cm} X}
\toprule
\textbf{Domain} & \textbf{Sub-axes} \\
\midrule
A. Macroeconomics & Multipliers (state dependence, non-linearities), growth/employment, debt/credibility, hysteresis, composition (spending vs. revenue), \textit{timing} (cycle, ZLB). \\
B. Social and Health & General/cause-specific mortality, morbidity, healthcare coverage and spending, poverty/inequity, gender and labor market, well-being/psychosocial. \\
C. Politics and Governance & Voting and electoral sanction, populism/attitudes, centralization, fiscal rules (EU), media and discourse, federalism/municipalism. \\
\bottomrule
\end{tabularx}
\end{table}

\paragraph*{Quality Assessment (Light Guideline 0--2)}
To weight interpretive strength without excluding evidence, a brief guideline (0–2) was applied across six dimensions (Table~\ref{tab:calidad}). The guideline informs the weight in the synthesis (not a publishable aggregate score).

\begin{table}[H]
\centering
\caption{Quality Assessment Guideline (0–2 per dimension)}
\label{tab:calidad}
\setlength{\tabcolsep}{6pt}
\renewcommand{\arraystretch}{1.12}
\begin{tabularx}{\textwidth}{L{4.8cm} L{7.9cm} L{2.1cm}}
\toprule
\textbf{Dimension} & \textbf{Criterion} & \textbf{Scale} \\
\midrule
Data transparency & Source, construction, access/DOI/repository & 0/1/2 \\
Identification/inference & Shock/IV/quasi-experiment/clear structure & 0/1/2 \\
External validity & Scope/translation and contextual limits & 0/1/2 \\
Replicability & Code, appendices, or sufficient detail & 0/1/2 \\
Pre-registration/ex-ante plan & Declaration or plan (if applicable) & 0/1/2 \\
Handling of biases & Publication/selection/measurement discussed & 0/1/2 \\
\bottomrule
\end{tabularx}
\end{table}

Synthesis proceeded in two stages: (i) \textit{descriptive syntax} (standardized summaries per study), and (ii) \textit{thematic integration} comparing direction and magnitude of effects by context (\textit{timing}, monetary constraint, sovereign risk) and composition of adjustment. Non-linearities (e.g., shocks $\geq$3\% of GDP) and state dependence (recession vs. expansion; ZLB vs. normality), as well as heterogeneity by region and period, were addressed.

To stress-test results, four checks were applied (Table~\ref{tab:robustez}). The narrative explicitly points out reasoned dissents and their methodological anchoring.

\begin{table}[H]
\centering
\caption{Robustness Checks Applied to Synthesis}
\label{tab:robustez}
\setlength{\tabcolsep}{6pt}
\renewcommand{\arraystretch}{1.12}
\begin{tabularx}{\textwidth}{L{5.2cm} X}
\toprule
\textbf{Check} & \textbf{Description} \\
\midrule
Sensitivity by design & Contrast of findings by method families (MRVAR vs. synthetic control vs. panel/QCA vs. ABA/AGENT-BASED). \\
Adjustment composition & Separation of spending vs. revenue; differential effects on public employment, transfers, and investment vs. taxes (income/consumption/capital). \\
Intensity and thresholds & Identification of non-linearities: acute shocks ($\geq 3\%$ GDP) vs. gradual consolidations. \\
Macro context & Stratification by monetary constraint (ZLB), capacity slack, and sovereign risk; verification of the ``expansionary'' hypothesis under high risk. \\
\bottomrule
\end{tabularx}
\end{table}

\paragraph*{Limitations}
(i) The \textit{narrative} character prevents quantitative meta-analysis; (ii) retrospective search reduces, but does not eliminate, risk of omission; (iii) heterogeneity of definitions of ``austerity'' among studies; (iv) light quality assessment, oriented towards weighting, not exclusion. These limitations are incorporated into the discussion when interpreting the strength and generality of conclusions.

\section{Macroeconomic Effects: Growth, Debt, and the Multiplier Debate}

One of the most intense debates in the literature centers on whether austerity is ``expansionary'' or, conversely, ``self-defeating'' \citep{dellepiane2014, muller2014}. The theory of ``expansionary fiscal consolidation'' (EFC) postulates that spending cuts, by signaling a commitment to fiscal discipline, can boost investor and consumer confidence, thus stimulating private investment and consumption, and offsetting the contraction in public demand \citep{botta2013b, nie2020}. However, an overwhelming majority of analyzed studies question this hypothesis, especially in the context of a recession. As noted by \citet{fontana2025}, EFC has ultimately proven to be an oxymoron in the European context.

Empirical evidence tends to support the Keynesian perspective that austerity has recessive effects, magnified by a fiscal multiplier that is larger during crises. \citet{jorda2013} use propensity score methods for time-series data and conclude that austerity is always a ``drag on growth,'' with a particularly severe impact in depressed economies. Their estimate is stark: a fiscal consolidation of 1\% of GDP leads to a reduction of 4\% in real GDP over a five-year period. Similarly, \citet{gechert2018}, analyzing the European case, find that fiscal multipliers were systematically underestimated by institutions, leading austerity measures to have persistent negative effects on potential output. These findings of ``hysteresis,'' where negative shocks have lasting effects, suggest that the short-term pain of austerity may not translate into long-term gains, but rather permanent output losses \citep{dosi2016, kleinmartins2024}. This idea is reinforced by \citet{libanio2020}, who, from a post-Keynesian perspective, argues that austerity policies can push the economy onto a new, lower growth trajectory.

The debate over the size of the fiscal multiplier is central to this discussion. Several studies point out that it is ``state-dependent'': it is larger during recessions, when monetary policy is constrained by the zero lower bound and there is idle capacity \citep{muller2014, semmler2013}. This implies that applying austerity at the worst moment of the business cycle is particularly damaging. In fact, \citet{blot2014} argue that delaying austerity could have improved growth prospects in the Eurozone without compromising debt reduction objectives. Research by \citet{rayl2020}, using the synthetic control method, quantifies the cost of austerity dramatically, estimating that, without it, GDP per capita in 2017 would have been 30\% higher in Greece, 11\% in Spain, and 17\% in Italy.

Nevertheless, the literature also introduces important nuances that complicate a universal condemnation of austerity in all circumstances. \citet{muller2014} maintain that while multipliers are large during financial crises, sovereign debt crises tend to reduce them. Therefore, when sovereign risk is high, austerity is ``unlikely to be literally self-defeating.'' Along similar lines, \citet{bianchi2019} and \citet{szkup2020} argue that in countries with high sovereign default risk, fiscal stimulus may be undesirable because it increases the probability of a debt crisis, theoretically justifying pro-cyclical policies. In this sense, austerity becomes a forced choice due to market pressures to maintain credibility \citep{mcmenamin2014}. However, \citet{mcmenamin2014}'s own study finds that, in practice, Eurozone bond markets did not respond positively to austere budgets; on the contrary, interest rates often increased, suggesting markets were skeptical about the efficacy of austerity in promoting growth and debt sustainability. Table \ref{tab:macro} summarizes a selection of key studies on macroeconomic effects, highlighting the tension between theory and empirical evidence.

\begin{center}
\begin{longtable}{@{}p{0.2\textwidth} p{0.4\textwidth} p{0.4\textwidth}@{}}
\caption{Synthesis of studies on the macroeconomic effects of austerity.} \label{tab:macro} \\
\toprule
\textbf{Author(s) and Year} & \textbf{Main Focus and Methodology} & \textbf{Key Findings} \\
\midrule
\endhead
\cite{muller2014} & Theoretical analysis on the fiscal multiplier in crises. & Multipliers are large in financial crises, but smaller in sovereign debt crises. Austerity is unlikely to be self-defeating when sovereign risk is high. \\
\cite{jorda2013} & Propensity score-based methods for time series to estimate the causal effect of fiscal policy. & Austerity is always a drag on growth, especially in depressed economies (a consolidation of 1\% of GDP reduces real GDP by 4\% in five years). \\
\cite{dosi2016} & Agent-based model (K+S) to compare Keynesian fiscal policies versus austerity rules. & Austerity increases volatility, unemployment, and deep crises. It reduces R\&D investment, is counterproductive, and increases the debt-to-GDP ratio. \\
\cite{gechert2018} & Analysis of fiscal forecast errors to estimate long-term effects of stimulus and austerity in Europe. & Early stimulus was beneficial in the long term, while subsequent austerity caused negative hysteresis effects and permanent output losses. \\
\cite{rayl2020} & Synthetic control method to evaluate GDP performance in Greece, Spain, and Italy post-2010. & Without fiscal consolidation, GDP per capita would have been significantly higher: 30\% in Greece, 11\% in Spain, and 17\% in Italy. \\
\cite{bilbaoubillos2014} & Regression analysis to examine the relationship between fiscal policies and growth in the Eurozone. & Fiscal multipliers were high and effects pro-cyclical during the crisis, confirming theoretical forecasts regarding the inefficacy of austerity in recession. \\
\cite{botta2013b} & Critical analysis of expansionary austerity theory. & Fiscal consolidation rarely has expansionary results, except in very specific and uncertain circumstances. \\
\cite{bianchi2019} & Model of endogenous sovereign default with nominal rigidities. & Increased public spending in recession reduces unemployment but increases the risk of a debt crisis, making stimulus undesirable in high-risk countries. \\
\cite{blot2014} & Simple model to analyze fiscal strategies and public debt trajectory simulations. & Delaying austerity could improve growth prospects without compromising debt reduction objectives, due to variation in multipliers over the cycle. \\
\cite{semmler2013} & Use of regime-switching models (MRVARs) to analyze fiscal consolidation in the EU. & Fiscal consolidation is state-dependent, being more contractive during recessions. The fiscal multiplier varies with economic conditions. \\
\cite{kleinmartins2024} & Analysis of non-linearities in fiscal multipliers using multiple datasets of fiscal shocks. & Austerity shocks exceeding 3\% of GDP have negative long-term effects on GDP, capital stock, and hours worked, persisting even after 15 years. \\
\bottomrule
\end{longtable}
\end{center}

\section{Social and Human Consequences: Health, Inequality, and Gender}

Beyond abstract macroeconomic debates, a substantial part of the literature has focused on the human costs of austerity. These studies consistently document how cuts in social spending, healthcare, education, and social safety nets have disproportionately affected the most vulnerable populations, exacerbating existing inequalities. As \citet{lusiani2018} state, ``austerity has a human face, and that face often belongs to the most marginalized members of society.''

In the realm of health, the evidence is particularly alarming. \citet{devogli2013} documents how strict austerity measures in countries like Greece, Spain, and Portugal correlated with an increase in suicides and outbreaks of infectious diseases, while Iceland, which rejected austerity, experienced minimal health impacts. \citet{kelsall2015} corroborate this view, arguing that austerity is a public health issue, as healthcare cuts and increased user fees deteriorate health outcomes. \citet{prante2020} apply this lesson to the COVID-19 crisis, showing how decades of austerity left the Italian healthcare system ill-prepared for the pandemic. However, not all studies are conclusive. \citet{saltkjel2017b}, using a fuzzy-set approach, found no consistent link between the combination of crisis and austerity and deteriorating health across Europe, suggesting a more complex relationship possibly mediated by other factors such as family support. Similarly, \citet{toffolutti2019} find that while austerity regimes are associated with a small increase in overall mortality, the results are sensitive to the countries included in the analysis. Austerity regimes are associated with an increase in all-cause mortality (0.7\%). At the same time, fiscal stimuli tend to significantly increase death rates from cirrhosis or chronic liver disease (3\%) and those due to road accidents (4.3\%). p. 1, our translation \citep{toffolutti2019}.

Austerity has also had profound consequences on income inequality and poverty. \citet{vegh2014} demonstrate that counter-cyclical fiscal policies (stimulus) reduce poverty, inequality, and unemployment, whereas pro-cyclical policies (austerity in recession) increase them, directly challenging the notion of ``expansionary fiscal austerity.'' \citet{bougrine2012} and \citet{stournaras2012} argue that austerity leads to an increase in poverty and inequality, posing a ``threat to socio-economic prosperity.'' From a human rights perspective, \citet{lusiani2018} maintain that austerity has undermined these rights and aggravated disparities based on income, gender, race, and disability, requiring the implementation of human rights impact assessments of fiscal consolidation policies.

The gender impact of austerity is another recurring theme. \citet{kushi2021} conclude that austerity policies during the Great Recession disproportionately harmed women in labor markets in the long term. Although men were initially more affected, cuts in the public sector and social services, where women are overrepresented, reversed this trend. \citet{perugini2018} confirm this finding for Europe, showing that austerity increased the gender wage gap and reduced women's chances of being employed in high-paying sectors. \citet{ortiz2011} specifically warn about the risks of public spending cuts for children and women in developing countries. Finally, the study by \citet{sambanis2022} on Greece offers a micro-social perspective, showing that job loss due to austerity decreased altruism and increased in-group bias in charitable donations, weakening social cohesion. Table \ref{tab:social} summarizes the findings of these studies.

\begin{center}
\begin{longtable}{@{}p{0.2\textwidth} p{0.4\textwidth} p{0.4\textwidth}@{}}
\caption{Synthesis of studies on the social consequences of austerity.} \label{tab:social} \\
\toprule
\textbf{Author(s) and Year} & \textbf{Main Focus and Methodology} & \textbf{Key Findings} \\
\midrule
\endhead
\cite{devogli2013} & Review of the health impact of austerity in Europe, comparing countries with strict policies (Greece, Spain) with Iceland. & Strict austerity was associated with an increase in suicides and outbreaks of infectious diseases. Iceland, which rejected austerity, had minimal health impacts. \\
\cite{kushi2021} & Case study of the Great Recession to analyze the gender effects of austerity on labor markets. & Austerity disproportionately harmed women in the long term by affecting sectors like services and public employment, and by cutting social safety nets. \\
\cite{vegh2014} & Analysis of four social indicators (poverty, inequality, unemployment, conflict) to examine the impact of fiscal policy responses in crises. & Counter-cyclical policy reduces poverty and inequality, while pro-cyclical policy (austerity in recession) increases them, challenging the idea of ``expansionary austerity.'' \\
\cite{perugini2018} & Analysis of EU-SILC data (2010-2013) and Blinder-Oaxaca decomposition to measure the impact of austerity on the gender pay gap. & Austerity increases gender wage inequality. Spending cuts have a greater effect than tax hikes. \\
\cite{ortiz2011} & Analysis of IMF spending projections in 128 developing countries to assess risks for children and women. & Widespread budget cuts since 2010, such as wage bill cuts and pension reforms, disproportionately affect children and women. \\
\cite{saltkjel2017b} & Fuzzy-set qualitative comparative analysis (fsQCA) with EU-SILC data for 29 European countries. & No consistent support was found for the thesis that the combination of crisis and austerity leads to a deterioration of health, suggesting a more complex relationship. \\
\cite{sambanis2022} & Household survey and modified dictator game in Greece to measure altruism after job loss. & Job loss is strongly associated with a decrease in altruism and an increase in in-group bias (preference for donating to national charities). \\
\cite{lusiani2018} & Development of a methodological framework for conducting Human Rights Impact Assessments (HRIAs) of fiscal consolidation. & Austerity has undermined human rights and exacerbated disparities, making HRIAs necessary to assess and mitigate these impacts. \\
\cite{toffolutti2019} & Analysis of data from 28 EU countries (1991-2013) to distinguish the effects of macroeconomic fluctuations and fiscal policies. & Austerity regimes are associated with a small increase (0.7\%) in overall mortality, but the results are sensitive to the sample of countries. \\
\cite{marston2014} & Qualitative case study on the impact of austerity on social welfare in Queensland, Australia. & Austerity measures were ineffective in reducing debt and promoting growth, but they increased state debt, unemployment, and eroded civil liberties. \\
\bottomrule
\end{longtable}
\end{center}

\section{Political and Governance Dynamics: Elections, Populism, and Institutional Power}

The implementation of austerity policies does not occur in a political vacuum. A third stream of literature has explored how austerity shapes and is shaped by politics, affecting electoral outcomes, the structure of governance, and public trust.

Several studies show that austerity is politically costly for incumbent governments. \citet{talving2018}, analyzing 24 European nations, finds that austerity measures lead to significantly lower levels of support for incumbents. Similarly, \citet{huebscher2018}, through surveys in five European countries, confirm that voters strongly oppose austerity, especially spending cuts. However, their study also reveals a paradox: despite this disapproval, they found no evidence that spending cuts are associated with greater leader turnover, suggesting that partisan loyalties may be stronger than opposition to a specific policy. \citet{fernandez2020} qualify this view, showing that public support for austerity can increase if specific reasons are made salient, although EU endorsement has no effect. Support is also divided by economic vulnerability and views on the euro. Conversely, \citet{bansak2019} and \citet{steelman2021} find, through surveys, that a majority of voters in several European countries do support austerity in principle, although this support is very sensitive to the specific design of the measures.

The relationship between austerity and the rise of populism is another key finding. \citet{crowley2020} argue that austerity policies since 2008 are linked to the rise of anti-immigrant attitudes and ethnic nationalism. Their analysis shows that areas most affected by fiscal austerity saw a corresponding increase in these attitudes. Even more compelling is the study by \citet{galofre2017a}, which uses historical data from Germany in the 1930s to show that areas most affected by fiscal austerity had a higher share of votes for the Nazi Party. This finding, robust across different methodological specifications, suggests a causal link between austerity-induced economic suffering and the appeal of extremist political solutions.

Austerity has also reshaped governance. In the context of crisis, governments tend to centralize decision-making. \citet{raudla2015}, based on a survey of public sector executives in 17 European countries, identify a ``centralization cascade,'' where the centralization of one element of the decision-making process leads to greater centralization throughout the system. \citet{herzog2019} add a layer of complexity by analyzing the internal politics of cabinets, showing that during economic crises, the prime minister's influence becomes more significant than that of the finance minister in budgetary allocations, weakening the mechanisms of fiscal authority delegation. At the EU level, the eurozone crisis and the austerity response led to the institutionalization of a centralized fiscal authority, known as the ``New Fiscal Union of Europe'' \citep{schlosser2019}, although its effectiveness remains a subject of debate \citep{franchino2019, katsikas2020}. Table \ref{tab:politica} summarizes these findings.

\begin{center}
\begin{longtable}{@{}p{0.2\textwidth} p{0.4\textwidth} p{0.4\textwidth}@{}}
\caption{Synthesis of studies on the political and governance dynamics of austerity.} \label{tab:politica} \\
\toprule
\textbf{Author(s) and Year} & \textbf{Main Focus and Methodology} & \textbf{Key Findings} \\
\midrule
\endhead
\cite{talving2018} & Analysis of economic policies as predictors of voting in 24 European nations (2004, 2009, 2014). & Fiscal austerity measures lead to significantly lower levels of support for incumbents. Responsibility lies primarily with national governments. \\
\cite{huebscher2018} & Survey experiments in five European countries and analysis of data from 32 countries since 1870. & Voters strongly oppose austerity, but there is no evidence that spending cuts increase leader turnover, possibly due to strong partisan loyalties. \\
\cite{galofre2017a} & Analysis of voting data in Germany (1930-1933) using an instrumental variable strategy and discontinuity design. & Areas with greater fiscal austerity had a relatively higher vote share for the Nazi Party. Austerity correlated with greater suffering and support for the Nazis. \\
\cite{crowley2020} & Regression analysis with data from UK and US electoral surveys. & Fiscal austerity policies since 2008 are linked to an increase in anti-immigrant and ethnic nationalist attitudes. \\
\cite{raudla2015} & Survey of public sector executives in 17 European countries to analyze the impact of the fiscal crisis on decision-making. & Fiscal crises trigger a ``centralization cascade,'' where centralization in one area leads to greater centralization throughout the government system. \\
\cite{afonso2014} & Comparative analysis of austerity reforms in Greece and Portugal, using documentary analysis and expert surveys. & Parties that rely on clientelistic links are more reluctant to accept austerity. The prevalence of these links in Greece generated conflicts, while in Portugal reforms were more consensual. \\
\cite{stanley2016} & Case study of the UK, analyzing the logic of anticipatory fiscal consolidation. & The governance of austerity in the UK is characterized by an anticipatory logic, focused on preventing future borrowing, which explains the slow pace of cuts. \\
\cite{herzog2019} & Analysis of contributions in budget debates in Ireland to estimate positions within the cabinet. & The delegation mechanism to the finance minister breaks down during crises, as the prime minister's influence becomes more significant in budget allocations. \\
\cite{bansak2019} & Original survey data and a conjoint experiment in five European countries. & A majority of voters prefer austerity to stimulus, but support depends on the specific design of the package and the party proposing it, with ideology being a key factor. \\
\cite{piferrer2019} & Content analysis of parliamentary debates in Portugal and Spain (1975-2017). & The term ``austerity'' became a global buzzword after 2008, spreading from fiscal policy to other areas, which facilitated its popularity. \\
\bottomrule
\end{longtable}
\end{center}

\section{Discussion: Convergences, Contradictions, and Research Gaps}

The synthesis of this extensive literature reveals several clear convergences. The first and most compelling is that austerity, especially when implemented during a recession, tends to be economically contractionary. The idea of an ``expansionary fiscal consolidation'' is, at best, a rare and conditional phenomenon, and at worst, a ``myth'' \citep{dellepiane2014, botta2016}. Multiple high-quality empirical studies, using various methodologies, confirm that fiscal multipliers are greater than one during crises, meaning that cuts in public spending reduce GDP by an amount greater than the cut itself \citep{jorda2013, gechert2018}.

A second convergence is the recognition of the profound social costs of austerity. There is a broad consensus that cuts in public services and social protection have exacerbated inequality, poverty, and labor precarity, with disproportionately negative impacts on vulnerable groups such as women, children, and minorities \citep{kushi2021, vegh2014, ortiz2011, perugini2018}. The connection between austerity and the deterioration of public health is also well-documented, although direct causality is more difficult to establish uniformly across all contexts \citep{devogli2013, toffolutti2019}.

The third area of agreement concerns the political consequences. Austerity is an unpopular policy that erodes support for incumbent governments \citep{talving2018} and is empirically linked to the rise of social discontent and support for populist and far-right parties \citep{crowley2020, galofre2017a}. This finding is crucial, as it suggests that economic policies cannot be separated from their implications for democratic stability.

Despite these convergences, the literature also presents contradictions and unresolved debates. The main contradiction remains the debate over whether austerity can be, under certain circumstances, a necessary or even beneficial policy. While most studies are critical, some suggest that in contexts of high sovereign risk and lack of access to debt markets, austerity may be the only option to avoid default, and that in such cases, its recessive effects may be smaller \citep{muller2014, bianchi2019}. This debate reflects a deeper tension between the economic logic of demand management and the logic of financial market confidence.

Another area of nuance is the design of austerity. Several studies distinguish between consolidations based on spending cuts and those based on tax increases. \citet{mcmanus2019} argue that tax increases on capital are the least harmful form of austerity, while cuts in transfers and public employment are the most regressive. \citet{jacques2020} finds that austerity tends to crowd out long-term investments in favor of politically more popular short-term programs, such as pensions. This suggests that not all austerity is the same, and its composition matters as much as its magnitude.

The review of this literature reveals some gaps. While there is a great deal of research on Europe and the United States, more analysis is needed on the effects of austerity in developing countries, where social safety nets are weaker and the human consequences may be even more severe \citep{ortiz2013}. Furthermore, although many studies document short- and medium-term effects, research on the very long-term consequences of austerity—for example, on human capital formation, innovation, and an economy's growth trajectory—is less developed, although works like those of \citet{dosi2016} and \citet{kleinmartins2024} are beginning to fill this gap. The COVID-19 crisis and subsequent fiscal responses open a new field of research to compare the lessons learned (or not learned) from the 2010 austerity era \citep{ban2020}.

\section{Conclusion}

This narrative review on fiscal austerity in times of crisis offers an unequivocal conclusion: austerity is a high-risk policy with deep, multifaceted, and negative consequences. Far from being a panacea for public debt, the evidence suggests that, when applied in the midst of a recession, it often proves to be macroeconomically counterproductive, exacerbating economic contraction and, in some cases, even worsening debt sustainability through its hysteresis effects on potential growth.

The social and human costs of austerity are equally significant and widely documented. The dismantling of public services, the erosion of social safety nets, and cuts in healthcare and education not only increase poverty and inequality but also disproportionately affect the most vulnerable groups in society. Studies on the impact on public health and gender inequality paint a grim picture of the human consequences of fiscal consolidation.

Politically, austerity has proven to be a catalyst for popular discontent. The loss of trust in institutions, the decline in support for traditional parties, and the alarming rise of populism and ethnic nationalism are linked, in part, to the economic and social suffering inflicted by these policies. The historical lesson that economic desperation can fuel political extremism resonates strongly in the findings of contemporary research.

While there are nuanced debates about the conditions under which austerity might be unavoidable or its effects less harmful—such as in cases of extreme sovereign debt crises—the weight of the evidence leans against its use as a universal tool for crisis management. The composition of austerity measures, their timing, and the institutional context are crucial in determining their outcomes.

Ultimately, the reviewed literature suggests that the question should not simply be ``austerity or stimulus?'' but rather, what kind of fiscal policy, at what time, and for whom? The lessons from the last decade of austerity call for a more pragmatic, equitable, and business cycle-sensitive approach to fiscal policy. As the world faces new crises, ignoring these lessons could mean repeating the same costly mistakes, with a high price for the economy, society, and democracy.


\begin{thebibliography}{}

\bibitem[Abels(2023)]{abels2023}
Abels, J. (2023). Does the current crisis mark the end of the EU’s austerity era? Competing political projects in European fiscal governance. \textit{Comparative European Politics, 22}(2), 192-211. \url{https://doi.org/10.1057/s41295-023-00346-4}

\bibitem[Afonso et al.(2014)]{afonso2014}
Afonso, A., Zartaloudis, S., \& Papadopoulos, Y. (2014). How party linkages shape austerity politics: clientelism and fiscal adjustment in Greece and Portugal during the eurozone crisis. \textit{Journal of European Public Policy, 22}(3), 315-334. \url{https://doi.org/10.1080/13501763.2014.964644}

\bibitem[Alvord(2019)]{alvord2019}
Alvord, D. (2019). The Triumph of Deficits: Supply-Side Economics, Institutional Constraints and the Political Articulation of Fiscal Crisis. \textit{The Sociological Quarterly, 61}(2), 206-230. \url{https://doi.org/10.1080/00380253.2019.1695552}

\bibitem[Arellano \& Bai(2016)]{arellano2016}
Arellano, C., \& Bai, Y. (2016). Fiscal Austerity during Debt Crises. \textit{Federal Reserve Bank of Minneapolis Staff Report, 525}. \url{https://doi.org/10.21034/sr.525}

\bibitem[Auerbach \& Gorodnichenko(2017)]{auerbach2017}
Auerbach, A., \& Gorodnichenko, Y. (2017). Fiscal Stimulus and Fiscal Sustainability. \textit{NBER Working Paper, No. w23789}. \url{https://doi.org/10.3386/w23789}

\bibitem[Ban(2020)]{ban2020}
Ban, C. (2020). Emergency Keynesianism 2.0: The political economy of fiscal policy in Europe during the Corona Crisis. \textit{Samfundsøkonomen, (4)}, 16-26. \url{https://doi.org/10.7146/samfundsokonomen.v0i4.123557}

\bibitem[Bandeira et al.(2021)]{bandeira2021}
Bandeira, G., Caballe Vilella, J., \& Vella, E. (2021). Emigration and Fiscal Austerity in a Depression. \textit{SSRN Electronic Journal}. \url{https://doi.org/10.2139/ssrn.3948391}

\bibitem[Bansak et al.(2019)]{bansak2019}
Bansak, K., Bechtel, M., \& Margalit, Y. (2019). Why Austerity? The Mass Politics of a Contested Policy. \textit{SSRN Electronic Journal}. \url{https://doi.org/10.2139/ssrn.3486227}

\bibitem[Bianchi et al.(2019)]{bianchi2019}
Bianchi, J., Ottonello, P., \& Presno, I. (2019). Fiscal Stimulus under Sovereign Risk. \textit{NBER Working Paper, No. w26307}. \url{https://doi.org/10.3386/w26307}

\bibitem[Bilbao-Ubillos \& Fernández-Sainz(2014)]{bilbaoubillos2014}
Bilbao-Ubillos, J., \& Fernández-Sainz, A. (2014). The impact of austerity policies in the Eurozone: fiscal multipliers and ‘adjustment fatigue’. \textit{Applied Economics Letters, 21}(14), 955-959. \url{https://doi.org/10.1080/13504851.2014.902013}

\bibitem[Bird \& Mandilaras(2013)]{bird2013}
Bird, G., \& Mandilaras, A. (2013). Fiscal imbalances and output crises in Europe: will the fiscal compact help or hinder?. \textit{Journal of Economic Policy Reform, 16}(1), 1-16. \url{https://doi.org/10.1080/17487870.2013.765081}

\bibitem[Blanco(2013)]{blanco2013}
Blanco, J. (2013). Review of Mark Blyth, Austerity. The History of a Dangerous Idea, New York: Oxford University Press, 2013. \textit{Journal of Philosophical Economics, 7}(1). \url{https://doi.org/10.46298/jpe.10657}

\bibitem[Blot et al.(2014)]{blot2014}
Blot, C., Cochard, M., Creel, J., Ducoudré, B., Schweisguth, D., \& Timbeau, X. (2014). Fiscal consolidation in times of crisis: is the sooner really the better?. \textit{Revue de l'OFCE, N° 132}(1), 159-192. \url{https://doi.org/10.3917/reof.132.0159}

\bibitem[Blueschke et al.(2016)]{blueschke2016}
Blueschke, D., Weyerstrass, K., \& Neck, R. (2016). How Should Slovenia Design Fiscal Policies in the Government Debt Crisis?. \textit{Emerging Markets Finance and Trade, 52}(7), 1562-1573. \url{https://doi.org/10.1080/1540496x.2016.1158549}

\bibitem[Blyth(2013)]{blyth2013}
Blyth, M. (2013). \textit{Austerity: The History of a Dangerous Idea}. Oxford University Press.

\bibitem[Botta(2012)]{botta2012}
Botta, A. (2012). Conflicting Claims in the Eurozone? Austerity's Myopic Logic and the Need for a European Federal Union in a Post-Keynesian Eurozone Center-Periphery Model. \textit{SSRN Electronic Journal}. \url{https://doi.org/10.2139/ssrn.2187410}

\bibitem[Botta(2013)]{botta2013b}
Botta, A. (2013). CARMEN REINHART AND KENNETH ROGOFF IN A TIME OF FISCAL AUSTERITY: A CRITICAL ANALYSIS OF THE EXPANSIONARY AUSTERITY THEORY. \textit{Istituto Lombardo - Accademia di Scienze e Lettere • Rendiconti di Lettere}. \url{https://doi.org/10.4081/let.2013.215}

\bibitem[Botta(2016)]{botta2016}
Botta, A. (2016). The Short- and Long-Run Inconsistency of the Expansionary Austerity Theory: A Post-Keynesian/Evolutionist Critique. \textit{SSRN Electronic Journal}. \url{https://doi.org/10.2139/ssrn.2884747}

\bibitem[Bougrine(2012)]{bougrine2012}
Bougrine, H. (2012). Fiscal austerity, the Great Recession and the rise of new dictatorships. \textit{Review of Keynesian Economics, 1}(0), 109-125. \url{https://doi.org/10.4337/roke.2012.01.07}

\bibitem[Boyer(2012)]{boyer2012}
Boyer, R. (2012). The four fallacies of contemporary austerity policies: the lost Keynesian legacy. \textit{Cambridge Journal of Economics, 36}(1), 283-312. \url{https://doi.org/10.1093/cje/ber037}

\bibitem[Castro \& Santos(2021)]{castro2021}
Castro, A., \& Santos, Í. (2021). Crises do Capital, Austeridade e Educação no Brasil. \textit{Research, Society and Development, 10}(2), e49810212523. \url{https://doi.org/10.33448/rsd-v10i2.12523}

\bibitem[Čekanavičius(2018)]{cekanavicius2018}
Čekanavičius, L. (2018). On the Choice of Fiscal Adjustment to Financial Crises: Expansionary vs. Contractionary Policies. \textit{Ekonomika, 97}(2), 7-17. \url{https://doi.org/10.15388/ekon.2018.1.11783}

\bibitem[Clift(2018a)]{clift2018a}
Clift, B. (2018a). The IMF, the eurozone and global financial crises, and the politics of economic ideas. \textit{Comparative European Politics, 18}(1), 99-108. \url{https://doi.org/10.1057/s41295-018-0146-x}

\bibitem[Clift(2018b)]{clift2018b}
Clift, B. (2018b). \textit{The IMF and the Politics of Austerity in the Wake of the Global Financial Crisis}. Oxford University Press. \url{https://doi.org/10.1093/oso/9780198813088.001.0001}

\bibitem[Cohen(2013)]{cohen2013}
Cohen, M. (2013). Austerity and the Global Crisis: Lessons from Latin America. \textit{Social Research: An International Quarterly, 80}(3), 929-952. \url{https://doi.org/10.1353/sor.2013.0073}

\bibitem[Crowley(2020)]{crowley2020}
Crowley, N. (2020). Austerity and ethno-nationalism. In \textit{Mapping Populism} (pp. 134-145). Routledge. \url{https://doi.org/10.4324/9780429295089-13}

\bibitem[Davidsson \& Bäck(2019)]{davidsson2019}
Davidsson, J., \& Bäck, H. (2019). Selecting Ministers in Times of Crisis: A Historical Analysis of the Role of Intra-Party Politics and Union Background in Swedish Cabinet Appointments 1917–2014. \textit{Political Studies, 67}(4), 932-954. \url{https://doi.org/10.1177/0032321718815847}

\bibitem[Davies(2021)]{davies2021}
Davies, J. (2021). Dynamics of Crisis, Neoliberalisation and Austerity. In \textit{Between Realism and Revolt} (pp. 47-70). Policy Press. \url{https://doi.org/10.1332/policypress/9781529210910.003.0003}

\bibitem[Dellepiane-Avellaneda(2014)]{dellepiane2014}
Dellepiane-Avellaneda, S. (2014). The Political Power of Economic Ideas: The Case of ‘Expansionary Fiscal Contractions'. \textit{The British Journal of Politics and International Relations, 17}(3), 391-418. \url{https://doi.org/10.1111/1467-856x.12038}

\bibitem[De Vogli(2013)]{devogli2013}
De Vogli, R. (2013). Financial crisis, austerity, and health in Europe. \textit{The Lancet, 382}(9890), 391. \url{https://doi.org/10.1016/s0140-6736(13)61662-1}

\bibitem[Di Mascio \& Natalini(2013)]{dimascio2013}
Di Mascio, F., \& Natalini, A. (2013). Fiscal Retrenchment in Southern Europe: Changing patterns of public management in Greece, Italy, Portugal and Spain. \textit{Public Management Review, 17}(1), 129-148. \url{https://doi.org/10.1080/14719037.2013.790275}

\bibitem[Díaz-Roldán et al.(2019)]{diazroldan2019}
Díaz-Roldán, C., Parada-Rodríguez, J., \& Carmona-González, N. (2019). Austerity policies in the Eurozone: How they affect youth unemployment?. \textit{The Central European Review of Economics and Management, 3}(2), 7. \url{https://doi.org/10.29015/cerem.753}

\bibitem[Díaz-Roldán(2017)]{diazroldan2017}
Díaz-Roldán, C. (2017). Fiscal performance in monetary unions: How much austerity should be allowed?. \textit{Panoeconomicus, 64}(1), 61-76. \url{https://doi.org/10.2298/pan140730021d}

\bibitem[Dosi et al.(2016)]{dosi2016}
Dosi, G., Napoletano, M., Roventini, A., \& Treibich, T. (2016). The Short- and Long-Run Damages of Fiscal Austerity: Keynes beyond Schumpeter. In \textit{Contemporary Issues in Macroeconomics} (pp. 79-100). Palgrave Macmillan. \url{https://doi.org/10.1057/9781137529589_9}

\bibitem[Dymski(2013)]{dymski2013}
Dymski, G. (2013). The Logic and Impossibility of Austerity. \textit{Social Research: An International Quarterly, 80}(3), 665-696. \url{https://doi.org/10.1353/sor.2013.0052}

\bibitem[Eaton(2017)]{eaton2017}
Eaton, D. (2017). Austerity and recovery in Ireland: Europe’s poster child and the great recession. \textit{Administration, 65}(4), 107-110. \url{https://doi.org/10.1515/admin-2017-0037}

\bibitem[Economides et al.(2021)]{economides2021}
Economides, G., Papageorgiou, D., \& Philippopoulos, A. (2021). Austerity, Assistance and Institutions: Lessons from the Greek Sovereign Debt Crisis. \textit{Open Economies Review, 32}(3), 435-478. \url{https://doi.org/10.1007/s11079-020-09613-3}

\bibitem[Edmiston(2014)]{edmiston2014}
Edmiston, D. (2014). The Age of Austerity: Contesting the Ethical Basis and Financial Sustainability of Welfare Reform in Europe. \textit{Journal of Contemporary European Studies, 22}(2), 118-131. \url{https://doi.org/10.1080/14782804.2014.910179}

\bibitem[Eggertsson(2014)]{eggertsson2014}
Eggertsson, G. (2014). Fiscal Policy, Public Debt and the World Crisis. \textit{German Economic Review, 15}(2), 225-242. \url{https://doi.org/10.1111/geer.12037}

\bibitem[Erber(2013)]{erber2013}
Erber, G. (2013). The Austerity Paradox: I See Austerity Everywhere, But Not in the Statistics. \textit{SSRN Electronic Journal}. \url{https://doi.org/10.2139/ssrn.2226319}

\bibitem[Fernández-Albertos \& Kuo(2016)]{fernandez2016}
Fernández-Albertos, J., \& Kuo, A. (2016). Economic Hardship and Policy Preferences in the Eurozone Periphery. \textit{Comparative Political Studies, 49}(7), 874-906. \url{https://doi.org/10.1177/0010414016633224}

\bibitem[Fernández-Albertos \& Kuo(2020)]{fernandez2020}
Fernandez-Albertos, J., \& Kuo, A. (2020). Selling Austerity: Preferences for Fiscal Adjustment during the Eurozone Crisis. \textit{Comparative Politics, 52}(2), 197-227. \url{https://doi.org/10.5129/001041520x15682460031849}

\bibitem[Fontana \& Sau(2025)]{fontana2025}
Fontana, O., \& Sau, L. (2025). Expansionary austerity in Europe: finally an oxymoron?. \textit{European Journal of Economics and Economic Policies: Intervention}, 1-26. \url{https://doi.org/10.4337/ejeep.2025.0152}

\bibitem[Franchino(2019)]{franchino2019}
Franchino, F. (2019). In search of the ideational foundations of EU fiscal governance: standard ideas, imperfect rules. \textit{Journal of European Integration, 42}(2), 179-194. \url{https://doi.org/10.1080/07036337.2019.1657858}

\bibitem[Galofré-Vilà et al.(2017)]{galofre2017a}
Galofré-Vilà, G., Meissner, C., McKee, M., \& Stuckler, D. (2017). Austerity and the Rise of the Nazi party. \textit{NBER Working Paper, No. w24106}. \url{https://doi.org/10.3386/w24106}

\bibitem[Gechert et al.(2018)]{gechert2018}
Gechert, S., Horn, G., \& Paetz, C. (2018). Long‐term Effects of Fiscal Stimulus and Austerity in Europe. \textit{Oxford Bulletin of Economics and Statistics, 81}(3), 647-666. \url{https://doi.org/10.1111/obes.12287}

\bibitem[Gournari(2016)]{gournari2016}
Gournari, P. (2016). The Necropolitics of Austerity: Discursive Constructions and Material Consequences in the Greek Context. \textit{Fast Capitalism, 13}(1). \url{https://doi.org/10.32855/fcapital.201601.004}

\bibitem[Groenendijk \& Jaansoo(2015)]{groenendijk2015}
Groenendijk, N., \& Jaansoo, A. (2015). Public Finance Systems for Coping with the Crises: Lessons from the Three Baltic States. \textit{Public Finance and Management, 15}(3), 257-289. \url{https://doi.org/10.1177/152397211501500305}

\bibitem[Győrffy(2015)]{gyorffy2015}
Győrffy, D. (2015). Austerity and growth in Central and Eastern Europe: understanding the link through contrasting crisis management in Hungary and Latvia. \textit{Post-Communist Economies, 27}(2), 129-152. \url{https://doi.org/10.1080/14631377.2015.1026682}

\bibitem[Hallerberg(2016)]{hallerberg2016}
Hallerberg, M. (2016). Fiscal Governance and Fiscal Outcomes Under EMU before and after the Crisis. In \textit{The Political and Economic Dynamics of the Eurozone Crisis} (pp. 145-166). Oxford University Press. \url{https://doi.org/10.1093/acprof:oso/9780198755739.003.0007}

\bibitem[Hannsgen(2012)]{hannsgen2012}
Hannsgen, G. (2012). Fiscal Policy, Unemployment Insurance, and Financial Crises in a Model of Growth And Distribution. \textit{SSRN Electronic Journal}. \url{https://doi.org/10.2139/ssrn.2061258}

\bibitem[Hardiman \& Dellepiane(2012)]{hardiman2012}
Hardiman, N., \& Dellepiane, S. (2012). The New Politics of Austerity: Fiscal Responses to Crisis in Ireland and Spain. \textit{SSRN Electronic Journal}. \url{https://doi.org/10.2139/ssrn.2013238}

\bibitem[Heald \& Hodges(2015)]{heald2015}
Heald, D., \& Hodges, R. (2015). Will “austerity” be a critical juncture in European public sector financial reporting?. \textit{Accounting, Auditing \& Accountability Journal, 28}(6), 993-1015. \url{https://doi.org/10.1108/aaaj-04-2014-1661}

\bibitem[Heiret \& Innset(2025)]{heiret2025}
Heiret, Y., \& Innset, O. (2025). Austerity without deficits: The global political economy of Norway’s fiscal paradox. \textit{Environment and Planning D: Society and Space}. \url{https://doi.org/10.1177/02637758241311020}

\bibitem[Helgason(2019)]{helgason2019}
Helgason, A. (2019). The Political Economy of Crisis Responses. In \textit{Welfare and the Great Recession} (pp. 43-58). Oxford University Press. \url{https://doi.org/10.1093/oso/9780198830962.003.0003}

\bibitem[Hellwig et al.(2020)]{hellwig2020}
Hellwig, T., Kweon, Y., \& Vowles, J. (2020). Shaping their own Destiny. In \textit{Democracy Under Siege?} (pp. 157-183). Oxford University Press. \url{https://doi.org/10.1093/oso/9780198846208.003.0007}

\bibitem[Herzog \& Jankin Mikhaylov(2019)]{herzog2019}
Herzog, A., \& Jankin Mikhaylov, S. (2019). Intra-cabinet politics and fiscal governance in times of austerity. \textit{Political Science Research and Methods, 8}(3), 409-424. \url{https://doi.org/10.1017/psrm.2019.40}

\bibitem[Hinkley(2015)]{hinkley2015}
Hinkley, S. (2015). Structurally adjusting: Narratives of fiscal crisis in four US cities. \textit{Urban Studies, 54}(9), 2123-2138. \url{https://doi.org/10.1177/0042098015618167}

\bibitem[Hodge(2019)]{hodge2019}
Hodge, B. (2019). Deep interdisciplinarity and responses to crisis. In \textit{Discourse Analysis and Austerity} (pp. 17-33). Routledge. \url{https://doi.org/10.4324/9781315208190-3}

\bibitem[Holgersen(2018)]{holgersen2018}
Holgersen, S. (2018). Searching for “Solutions” to Crisis: A Critique of Urban Austerity and Keynesianism. \textit{Human Geography, 11}(2), 38-58. \url{https://doi.org/10.1177/194277861801100204}

\bibitem[Huebscher et al.(2018)]{huebscher2018}
Huebscher, E., Sattler, T., \& Wagner, M. (2018). Voter Responses to Fiscal Austerity. \textit{SSRN Electronic Journal}. \url{https://doi.org/10.2139/ssrn.3289341}

\bibitem[Jacques(2020)]{jacques2020}
Jacques, O. (2020). Austerity and the path of least resistance: how fiscal consolidations crowd out long-term investments. \textit{Journal of European Public Policy, 28}(4), 551-570. \url{https://doi.org/10.1080/13501763.2020.1737957}

\bibitem[Jordà \& Taylor(2013)]{jorda2013}
Jordà, Ò., \& Taylor, A. (2013). The Time for Austerity: Estimating the Average Treatment Effect of Fiscal Policy. \textit{NBER Working Paper, No. w19414}. \url{https://doi.org/10.3386/w19414}

\bibitem[Karger(2014)]{karger2014}
Karger, H. (2014). The Bitter Pill: Austerity, Debt, and the Attack on Europe's Welfare States. \textit{The Journal of Sociology \& Social Welfare, 41}(2). \url{https://doi.org/10.15453/0191-5096.3949}

\bibitem[Katsikas(2020)]{katsikas2020}
Katsikas, D. (2020). Fiscal Governance in the Eurozone: From Maastricht to crisis and back again?*. \textit{Region \& Periphery, (9)}, 83. \url{https://doi.org/10.12681/rp.23780}

\bibitem[Kelsall \& Hennings(2015)]{kelsall2015}
Kelsall, D., \& Hennings, C. (2015). Austerity and fiscal prudence are health issues. \textit{Canadian Medical Association Journal, 187}(14), 1029. \url{https://doi.org/10.1503/cmaj.150950}

\bibitem[Kim \& Warner(2020)]{kim2020}
Kim, Y., \& Warner, M. (2020). Pragmatic municipalism or austerity urbanism? Understanding local government responses to fiscal stress. \textit{Local Government Studies, 47}(2), 234-252. \url{https://doi.org/10.1080/03003930.2020.1729751}

\bibitem[Kinsella(2012)]{kinsella2012}
Kinsella, S. (2012). Is Ireland really the role model for austerity?. \textit{Cambridge Journal of Economics, 36}(1), 223-235. \url{https://doi.org/10.1093/cje/ber032}

\bibitem[Kitromilides(2011)]{kitromilides2011}
Kitromilides, Y. (2011). Deficit reduction, the age of austerity, and the paradox of insolvency. \textit{Journal of Post Keynesian Economics, 33}(3), 517-536. \url{https://doi.org/10.2753/pke0160-3477330306}

\bibitem[Klein Martins(2024)]{kleinmartins2024}
Klein Martins, G. (2024). Long‐run Effects of Austerity: An Analysis of Size Dependence and Persistence in Fiscal Multipliers. \textit{Oxford Bulletin of Economics and Statistics, 87}(2), 330-356. \url{https://doi.org/10.1111/obes.12646}

\bibitem[Kramer(2019)]{kramer2019}
Kramer, Z. (2019). Fiscal Sovereignty under EU Crisis Management: A Comparison of Greece and Hungary. \textit{Acta Oeconomica, 69}(4), 595-624. \url{https://doi.org/10.1556/032.2019.69.4.6}

\bibitem[Kuang \& Mitra(2024)]{kuang2024}
Kuang, P., \& Mitra, K. (2024). Potential Output Pessimism and Austerity in the European Union. \textit{SSRN Electronic Journal}. \url{https://doi.org/10.2139/ssrn.4915488}

\bibitem[Kushi \& McManus(2021)]{kushi2021}
Kushi, S., \& McManus, I. (2021). Gender, austerity and the welfare state. In \textit{Handbook on Austerity, Populism and the Welfare State}. Edward Elgar Publishing. \url{https://doi.org/10.4337/9781789906745.00033}

\bibitem[Labonté(2012)]{labonte2012}
Labonté, R. (2012). The austerity agenda: how did we get here and where do we go next?. \textit{Critical Public Health, 22}(3), 257-265. \url{https://doi.org/10.1080/09581596.2012.687508}

\bibitem[Larson(2012)]{larson2012}
Larson, S. R. (2012). Austerity: Causes, Consequences and Remedies. \textit{SSRN Electronic Journal}. \url{https://doi.org/10.2139/ssrn.2048739}

\bibitem[Lebaron(2018)]{lebaron2018}
Lebaron, F. (2018). Sociologia e ciências sociais em tempos de austeridade. \textit{Sociedade e Estado, 33}(2), 529-537. \url{https://doi.org/10.1590/s0102-699220183302012}

\bibitem[Leblond(2013)]{leblond2013}
Leblond, P. (2013). Fiscal Crises in the Eurozone: Assessing the Austerity Imposed by the Bailouts. In \textit{Economic Crisis in Europe} (pp. 67-86). Palgrave Macmillan. \url{https://doi.org/10.1057/9781137005236_4}

\bibitem[Libanio(2020)]{libanio2020}
Libanio, G. (2020). CARDIM DE CARVALHO AND THE POST KEYNESIANS ON FISCAL POLICY: THE ECONOMIC CONSEQUENCES OF AUSTERITY. \textit{Revista de Economia Contemporânea, 24}(2). \url{https://doi.org/10.1590/198055272429}

\bibitem[Lusiani \& Chaparro(2018)]{lusiani2018}
Lusiani, N., \& Chaparro, S. (2018). Assessing Austerity: Monitoring the Human Rights Impacts of Fiscal Consolidation. \textit{SSRN Electronic Journal}. \url{https://doi.org/10.2139/ssrn.3218609}

\bibitem[Mabbett \& Schelkle(2014)]{mabbett2014}
Mabbett, D., \& Schelkle, W. (2014). Searching Under the Lamp-Post: The Evolution of Fiscal Surveillance. \textit{SSRN Electronic Journal}. \url{https://doi.org/10.2139/ssrn.2434008}

\bibitem[Malin(2020)]{malin2020}
Malin, N. (2020). Austerity as a UK policy context in the early twenty-first century. In \textit{De-Professionalism and Austerity} (pp. 3-12). Policy Press. \url{https://doi.org/10.1332/policypress/9781447350163.003.0001}

\bibitem[Markantonatou(2013)]{markantonatou2013}
Markantonatou, M. (2013). Fiscal Discipline Through Internal Devaluation and Discourses of Rent-Seeking: The Case of the Crisis in Greece. \textit{Studies in Political Economy, 91}(1), 59-83. \url{https://doi.org/10.1080/19187033.2013.11674982}

\bibitem[Marston(2014)]{marston2014}
Marston, G. (2014). Queensland's Budget Austerity and Its Impact on Social Welfare: Is the Cure Worse than the Disease?. \textit{The Journal of Sociology \& Social Welfare, 41}(2). \url{https://doi.org/10.15453/0191-5096.3955}

\bibitem[Mastromatteo \& Rossi(2015)]{mastromatteo2015}
Mastromatteo, G., \& Rossi, S. (2015). The economics of deflation in the euro area: a critique of fiscal austerity. \textit{Review of Keynesian Economics, 3}(3), 336-350. \url{https://doi.org/10.4337/roke.2015.03.04}

\bibitem[McBride \& Merolli(2013)]{mcbride2013}
McBride, S., \& Merolli, J. (2013). Alternatives to austerity? Post-crisis policy advice from global institutions. \textit{Global Social Policy, 13}(3), 299-320. \url{https://doi.org/10.1177/1468018113499980}

\bibitem[McManus et al.(2019)]{mcmanus2019}
McManus, R., Ozkan, F., \& Trzeciakiewicz, D. (2019). Fiscal consolidations and distributional effects: which form of fiscal austerity is least harmful?. \textit{Oxford Economic Papers, 73}(1), 317-349. \url{https://doi.org/10.1093/oep/gpz065}

\bibitem[McManus(2015)]{mcmanus2015}
McManus, R. (2015). Austerity versus stimulus: the polarizing effect of fiscal policy. \textit{Oxford Economic Papers, 67}(3), 581-597. \url{https://doi.org/10.1093/oep/gpv023}

\bibitem[McMenamin et al.(2014)]{mcmenamin2014}
McMenamin, I., Breen, M., \& Muñoz-Portillo, J. (2014). Austerity and credibility in the Eurozone. \textit{European Union Politics, 16}(1), 45-66. \url{https://doi.org/10.1177/1465116514553487}

\bibitem[Mercille(2013)]{mercille2013}
Mercille, J. (2013). The role of the media in fiscal consolidation programmes: the case of Ireland. \textit{Cambridge Journal of Economics, 38}(2), 281-300. \url{https://doi.org/10.1093/cje/bet068}

\bibitem[Montgomerie \& Tepe-Belfrage(2020)]{montgomerie2020}
Montgomerie, J., \& Tepe-Belfrage, D. (2020). Financialisation, crisis and austerity as the distribution of harm. In \textit{The Routledge Handbook of Critical European Studies} (pp. 201-211). Routledge. \url{https://doi.org/10.4324/9780429491306-14}

\bibitem[Müller(2014)]{muller2014}
Müller, G. (2014). Fiscal Austerity and the Multiplier in Times of Crisis. \textit{German Economic Review, 15}(2), 243-258. \url{https://doi.org/10.1111/geer.12027}

\bibitem[Nahtigal \& Bugaric(2012)]{nahtigal2012}
Nahtigal, M., \& Bugaric, B. (2012). The EU Fiscal Compact: Constitutionalization of Austerity and Preemption of Democracy in Europe. \textit{SSRN Electronic Journal}. \url{https://doi.org/10.2139/ssrn.2194475}

\bibitem[Neck et al.(2013)]{neck2013}
Neck, R., Blueschke, D., \& Weyerstrass, K. (2013). Trade-Off of Fiscal Austerity in the European Debt Crisis in Slovenia. \textit{International Advances in Economic Research, 19}(4), 367-380. \url{https://doi.org/10.1007/s11294-013-9438-8}

\bibitem[Nie(2020)]{nie2020}
Nie, O. (2020). \textit{Expansionary Fiscal Austerity: New International Evidence}. World Bank. \url{https://doi.org/10.1596/1813-9450-9344}

\bibitem[Nikiforos(2021)]{nikiforos2021}
Nikiforos, M. (2021). Crisis, austerity, and fiscal expenditure in Greece: recent experience and future prospects in the post-COVID-19 era. \textit{European Journal of Economics and Economic Policies: Intervention}, 1-18. \url{https://doi.org/10.4337/ejeep.2021.0076}

\bibitem[Orair \& Gobetti(2017)]{orair2017}
Orair, R., \& Gobetti, S. (2017). Brazilian Fiscal Policy in Perspective: From Expansion to Austerity. In \textit{The Brazilian Economy since the Great Financial Crisis of 2007/2008} (pp. 219-244). Springer. \url{https://doi.org/10.1007/978-3-319-64885-9_9}

\bibitem[Ortiz \& Cummins(2013)]{ortiz2013}
Ortiz, I., \& Cummins, M. (2013). The Age of Austerity: A Review of Public Expenditures and Adjustment Measures in 181 Countries. \textit{SSRN Electronic Journal}. \url{https://doi.org/10.2139/ssrn.2260771}

\bibitem[Ortiz \& Cummins(2019)]{ortiz2019}
Ortiz, I., \& Cummins, M. (2019). Austerity: The New Normal - A Renewed Washington Consensus 2010-24. \textit{SSRN Electronic Journal}. \url{https://doi.org/10.2139/ssrn.3523562}

\bibitem[Ortiz \& Cummins(2020)]{ortiz2020}
Ortiz, I., \& Cummins, M. (2020). The Austerity Decade 2010-20. \textit{Social Policy and Society, 20}(1), 142-157. \url{https://doi.org/10.1017/s1474746420000433}

\bibitem[Ortiz \& Cummins(2021)]{ortiz2021}
Ortiz, I., \& Cummins, M. (2021). Global Austerity Alert: Looming Budget Cuts in 2021-25 and Alternative Pathways. \textit{SSRN Electronic Journal}. \url{https://doi.org/10.2139/ssrn.3856299}

\bibitem[Ortiz et al.(2011)]{ortiz2011}
Ortiz, I., Chai, J., \& Cummins, M. (2011). Austerity Measures Threaten Children and Poor Households: Recent Evidence in Public Expenditures from 128 Developing Countries. \textit{SSRN Electronic Journal}. \url{https://doi.org/10.2139/ssrn.1934510}

\bibitem[Overmans \& Noordegraaf(2014)]{overmans2014}
Overmans, J., \& Noordegraaf, M. (2014). Managing austerity: rhetorical and real responses to fiscal stress in local government. \textit{Public Money \& Management, 34}(2), 99-106. \url{https://doi.org/10.1080/09540962.2014.887517}

\bibitem[Papadia(2023)]{papadia2023}
Papadia, A. (2023). Fiscal policy under constraints: Fiscal capacity and austerity during the Great Depression. \textit{The Economic History Review, 77}(1), 90-118. \url{https://doi.org/10.1111/ehr.13253}

\bibitem[Parguez(1989)]{parguez1989}
Parguez, A. (1989). Cet âge de l’austérité. \textit{Économie appliquée, 42}(1), 71-89. \url{https://doi.org/10.3406/ecoap.1989.2117}

\bibitem[Peck(2013)]{peck2013}
Peck, J. (2013). Pushing austerity: state failure, municipal bankruptcy and the crises of fiscal federalism in the USA. \textit{Cambridge Journal of Regions, Economy and Society, 7}(1), 17-44. \url{https://doi.org/10.1093/cjres/rst018}

\bibitem[Peck(2020)]{peck2020}
Peck, J. (2020). Austerity Urbanism. In \textit{Oxford Bibliographies in Urban Studies}. Oxford University Press. \url{https://doi.org/10.1093/obo/9780190922481-0001}

\bibitem[Perugini et al.(2018)]{perugini2018}
Perugini, C., Žarković Rakić, J., \& Vladisavljević, M. (2018). Austerity and gender inequalities in Europe in times of crisis. \textit{Cambridge Journal of Economics, 43}(3), 733-767. \url{https://doi.org/10.1093/cje/bey044}

\bibitem[Peters(2011)]{peters2011}
Peters, B. (2011). Governance responses to the fiscal crisis—comparative perspectives. \textit{Public Money \& Management, 31}(1), 75-80. \url{https://doi.org/10.1080/09540962.2011.545551}

\bibitem[Petkov(2014)]{petkov2014}
Petkov, V. (2014). Advantages and Disadvantages of Fiscal Discipline in Bulgaria in Times of Crisis. \textit{Contemporary Economics, 8}(1), 47-56. \url{https://doi.org/10.5709/ce.1897-9254.130}

\bibitem[Petzold(2022)]{petzold2022}
Petzold, T. (2022). Normalisierte Austerität in der BRD. \textit{Geographische Zeitschrift, 110}(3), 127. \url{https://doi.org/10.25162/gz-2022-0004}

\bibitem[Pi Ferrer \& Alasuutari(2019)]{piferrer2019}
Pi Ferrer, L., \& Alasuutari, P. (2019). The Spread and Domestication of the Term “Austerity:” Evidence from the Portuguese and Spanish Parliaments. \textit{Politics \& Policy, 47}(6), 1039-1065. \url{https://doi.org/10.1111/polp.12331}

\bibitem[Pitty(n.d.)]{pitty_nd}
Pitty, R. (n.d.). Disintegrating European Austerity in Greece and Germany. In \textit{Democracy and Crisis}. Palgrave Macmillan. \url{https://doi.org/10.1057/9781137326041.0012}

\bibitem[Posner \& Sommerfeld(2012)]{posner2012}
Posner, P. L., \& Sommerfeld, M. (2012). The Politics of Fiscal Austerity: Implications for the United States. \textit{Public Budgeting \& Finance, 32}(3), 32-52. \url{https://doi.org/10.1111/j.1540-5850.2012.01020.x}

\bibitem[Prante et al.(2020)]{prante2020}
Prante, F., Bramucci, A., \& Truger, A. (2020). Decades of Tight Fiscal Policy Have Left the Health Care System in Italy Ill-Prepared to Fight the COVID-19 Outbreak. \textit{Intereconomics, 55}(3), 147-152. \url{https://doi.org/10.1007/s10272-020-0886-0}

\bibitem[Prota \& Grisorio(2018)]{prota2018}
Prota, F., \& Grisorio, M. (2018). Public expenditure in time of crisis: are Italian policymakers choosing the right mix?. \textit{Economia Politica, 35}(2), 337-365. \url{https://doi.org/10.1007/s40888-018-0124-4}

\bibitem[Pühringer(2019)]{puhringer2019}
Pühringer, S. (2019). The ‘eternal character’ of austerity measures in European crisis policies: Evidence from the Fiscal Compact discourse in Austria. In \textit{Discourse Analysis and Austerity} (pp. 237-252). Routledge. \url{https://doi.org/10.4324/9781315208190-17}

\bibitem[Raess(2021)]{raess2021}
Raess, D. (2021). Globalization and Austerity: Flipping Partisan Effects on Fiscal Policy During (Recent) International Crises. \textit{Political Studies, 71}(2), 332-358. \url{https://doi.org/10.1177/00323217211015811}

\bibitem[Randma-Liiv \& Kickert(2017)]{randmaliiv2017}
Randma-Liiv, T., \& Kickert, W. (2017). The Impact of Fiscal Crisis on Public Administration in Europe. In \textit{The Palgrave Handbook of Public Administration and Management in Europe} (pp. 899-917). Palgrave Macmillan. \url{https://doi.org/10.1057/978-1-137-55269-3_47}

\bibitem[Raudla \& Bur(2022)]{raudla2022}
Raudla, R., \& Bur, S. (2022). Austerity and the use of performance information in the budget process. \textit{Public Money \& Management, 43}(6), 627-634. \url{https://doi.org/10.1080/09540962.2021.2020976}

\bibitem[Raudla et al.(2015)]{raudla2015}
Raudla, R., Douglas, J., Randma‐Liiv, T., \& Savi, R. (2015). The Impact of Fiscal Crisis on Decision‐Making Processes in European Governments: Dynamics of a Centralization Cascade. \textit{Public Administration Review, 75}(6), 842-852. \url{https://doi.org/10.1111/puar.12381}

\bibitem[Raudla \& Tavares(2017)]{raudla2017}
Raudla, R., \& Tavares, A. (2017). Inter-municipal Cooperation and Austerity Policies: Obstacles or Opportunities?. In \textit{Governance and Public Management} (pp. 17-41). Springer. \url{https://doi.org/10.1007/978-3-319-62819-6_2}

\bibitem[Raudla(2011)]{raudla2011}
Raudla, R. (2011). FISCAL RETRENCHMENT IN ESTONIA DURING THE FINANCIAL CRISIS: THE ROLE OF INSTITUTIONAL FACTORS. \textit{Public Administration, 91}(1), 32-50. \url{https://doi.org/10.1111/j.1467-9299.2011.01963.x}

\bibitem[Raudon \& Shore(2018)]{raudon2018}
Raudon, S., \& Shore, C. (2018). The Eurozone Crisis, Greece and European Integration. \textit{Anthropological Journal of European Cultures, 27}(1), 64-83. \url{https://doi.org/10.3167/ajec.2018.270111}

\bibitem[Rayl(2020)]{rayl2020}
Rayl, N. (2020). Cost of Austerity: Effect of Fiscal Consolidation in Europe Post 2010. \textit{SSRN Electronic Journal}. \url{https://doi.org/10.2139/ssrn.3596470}

\bibitem[Robbins \& Lapsley(2014)]{robbins2014}
Robbins, G., \& Lapsley, I. (2014). The success story of the Eurozone crisis? Ireland's austerity measures. \textit{Public Money \& Management, 34}(2), 91-98. \url{https://doi.org/10.1080/09540962.2014.887515}

\bibitem[Russell(2019)]{russell2019}
Russell, E. D. (2019). Austerity and the eclipse of economic alternatives. In \textit{Discourse Analysis and Austerity} (pp. 34-49). Routledge. \url{https://doi.org/10.4324/9781315208190-4}

\bibitem[Saldaña(2024)]{saldana2024}
Saldaña, C. M. (2024). Accountability or Austerity? Examining the Practice of K–12 Early Fiscal Intervention During Periods of Economic Crisis. \textit{Educational Evaluation and Policy Analysis, 47}(3), 907-938. \url{https://doi.org/10.3102/01623737241254841}

\bibitem[Saltkjel et al.(2017a)]{saltkjel2017a}
Saltkjel, T., Ingelsrud, M., Dahl, E., \& Halvorsen, K. (2017a). A fuzzy set approach to economic crisis, austerity and public health. Part I. European countries’ conformity to ideal types during the economic downturn. \textit{Scandinavian Journal of Public Health, 45}(18\_suppl), 41-47. \url{https://doi.org/10.1177/1403494817706632}

\bibitem[Saltkjel et al.(2017b)]{saltkjel2017b}
Saltkjel, T., Holm Ingelsrud, M., Dahl, E., \& Halvorsen, K. (2017b). A fuzzy set approach to economic crisis, austerity and public health. Part II: How are configurations of crisis and austerity related to changes in population health across Europe?. \textit{Scandinavian Journal of Public Health, 45}(18\_suppl), 48-55. \url{https://doi.org/10.1177/1403494817707125}

\bibitem[Sambanis et al.(2022)]{sambanis2022}
Sambanis, N., Nikolova, E., \& Schultz, A. (2022). The effects of economic austerity on pro-sociality: Evidence from Greece. \textit{European Union Politics, 23}(4), 567-589. \url{https://doi.org/10.1177/14651165221120527}

\bibitem[Sawyer(2012)]{sawyer2012}
Sawyer, M. (2012). The tragedy of UK fiscal policy in the aftermath of the financial crisis. \textit{Cambridge Journal of Economics, 36}(1), 205-221. \url{https://doi.org/10.1093/cje/ber043}

\bibitem[Schlosser(2019)]{schlosser2019}
Schlosser, P. (2019). \textit{Europe's New Fiscal Union}. Springer. \url{https://doi.org/10.1007/978-3-319-98636-4}

\bibitem[Schuknecht(2010)]{schuknecht2010}
Schuknecht, L. (2010). Fiscal Activism in Booms, Busts, and Beyond. \textit{SSRN Electronic Journal}. \url{https://doi.org/10.2139/ssrn.1985185}

\bibitem[Seccareccia(2012)]{seccareccia2012}
Seccareccia, M. (2012). Understanding Fiscal Policy and the New Fiscalism. \textit{International Journal of Political Economy, 41}(2), 61-81. \url{https://doi.org/10.2753/ijp0891-1916410204}

\bibitem[Seidman(2012)]{seidman2012}
Seidman, L. (2012). Keynesian stimulus versus classical austerity. \textit{Review of Keynesian Economics, 1}(0), 77-92. \url{https://doi.org/10.4337/roke.2012.01.05}

\bibitem[Semmler \& Semmler(2013)]{semmler2013}
Semmler, W., \& Semmler, A. (2013). The Macroeconomics of Fiscal Consolidation in the European Union. \textit{SSRN Electronic Journal}. \url{https://doi.org/10.2139/ssrn.2320198}

\bibitem[Shapiro(2012)]{shapiro2012}
Shapiro, N. (2012). Keynes, Steindl, and the Critique of Austerity Economics. \textit{Monthly Review, 64}(3), 103. \url{https://doi.org/10.14452/mr-064-03-2012-07_7}

\bibitem[Silva(2016)]{silva2016}
Silva, C. (2016). \textit{Fiscal Austerity and Innovation in Local Governance in Europe}. Routledge. \url{https://doi.org/10.4324/9781315582474}

\bibitem[Silva(2023)]{silva2023}
Silva, D. H. (2023). Optimal Fiscal Consolidation Under Frictional Financial Markets. \textit{The Economic Journal, 133}(652), 1537-1585. \url{https://doi.org/10.1093/ej/uead013}

\bibitem[Škare \& Druzeta(2015)]{skare2015}
Škare, M., \& Druzeta, R. P. (2015). Fiscal Austerity Versus Growth in Croatia. \textit{Contemporary Economics, 9}(1), 77-92. \url{https://doi.org/10.5709/ce.1897-9254.162}

\bibitem[Stanley(2016)]{stanley2016}
Stanley, L. (2016). Governing austerity in the United Kingdom: anticipatory fiscal consolidation as a variety of austerity governance. \textit{Economy and Society, 45}(3-4), 303-324. \url{https://doi.org/10.1080/03085147.2016.1224145}

\bibitem[Steelman(2021)]{steelman2021}
Steelman, T. (2021). Do Voters Support Austerity Measures in Times of Economic Crisis? For Many Voters, the Answer is Yes. \textit{Political Science Today, 1}(3), 7. \url{https://doi.org/10.1017/psj.2021.52}

\bibitem[Stournaras(2012)]{stournaras2012}
Stournaras, C. F. (2012). Fiscal Austerity as the Achilles’ Heel to Socio-Economic Prosperity. \textit{SSRN Electronic Journal}. \url{https://doi.org/10.2139/ssrn.2111896}

\bibitem[Szkup(2020)]{szkup2020}
Szkup, M. (2020). Preventing Self-fulfilling Debt Crises: A Global Games Approach. \textit{SSRN Electronic Journal}. \url{https://doi.org/10.2139/ssrn.3527210}

\bibitem[Talving(2018)]{talving2018}
Talving, L. (2018). The electoral consequences of austerity: economic policy voting in Europe in times of crisis. In \textit{Electoral Rules and Electoral Behaviour} (pp. 58-81). Routledge. \url{https://doi.org/10.4324/9781351273527-4}

\bibitem[Theodore(2019)]{theodore2019}
Theodore, N. (2019). Governing through austerity: (Il)logics of neoliberal urbanism after the global financial crisis. \textit{Journal of Urban Affairs, 42}(1), 1-17. \url{https://doi.org/10.1080/07352166.2019.1623683}

\bibitem[Toffolutti \& Suhrcke(2019)]{toffolutti2019}
Toffolutti, V., \& Suhrcke, M. (2019). Does austerity really kill?. \textit{SocArXiv}. \url{https://doi.org/10.31235/osf.io/b2t4x}

\bibitem[Toschi et al.(2024)]{toschi2024}
Toschi, B., Mendes, Á. N., \& Carnut, L. (2024). The Debate on the Fiscal Matter in the Contemporary Capitalism Crisis in Light of Critical Political Economy. \textit{Modern Economy, 15}(01), 1-22. \url{https://doi.org/10.4236/me.2024.151001}

\bibitem[Truger(2013)]{truger2013}
Truger, A. (2013). Austerity in the euro area: the sad state of economic policy in Germany and the EU. \textit{European Journal of Economics and Economic Policies: Intervention, 10}(2), 158-174. \url{https://doi.org/10.4337/ejeep.2013.02.02}

\bibitem[Ufodike(2020)]{ufodike2020}
Ufodike, A. (2020). Reviving Austerity: Populist Support for Unpopular Economics in Canada. \textit{SSRN Electronic Journal}. \url{https://doi.org/10.2139/ssrn.3720355}

\bibitem[Veebel \& Kulu(2014)]{veebel2014}
Veebel, V., \& Kulu, L. (2014). Against the political expectations and theoretical models: how to implement austerity and not to lose political power. \textit{Baltic Journal of Economics, 14}(1-2), 2-16. \url{https://doi.org/10.1080/1406099x.2014.991140}

\bibitem[Vegh \& Vuletin(2014)]{vegh2014}
Vegh, C., \& Vuletin, G. (2014). Social Implications of Fiscal Policy Responses During Crises. \textit{NBER Working Paper, No. w19828}. \url{https://doi.org/10.3386/w19828}

\bibitem[Warwick(2018)]{warwick2018}
Warwick, B. T. C. (2018). Debt, Austerity, and the Structural Responses of Social Rights. In \textit{Sovereign Debt and Human Rights} (pp. 381-401). Oxford University Press. \url{https://doi.org/10.1093/oso/9780198810445.003.0021}

\bibitem[Watts \& Sharpe(2013)]{watts2013}
Watts, M. J., \& Sharpe, T. P. (2013). Immutable laws of debt dynamics. \textit{Journal of Post Keynesian Economics, 36}(1), 59-84. \url{https://doi.org/10.2753/pke0160-3477360104}

\bibitem[Zahariadis(2013)]{zahariadis2013}
Zahariadis, N. (2013). The Politics of Risk-sharing: Fiscal Federalism and the Greek Debt Crisis. \textit{Journal of European Integration, 35}(3), 271-285. \url{https://doi.org/10.1080/07036337.2013.774787}

\bibitem[Zezza(2012)]{zezza2012}
Zezza, G. (2012). The impact of fiscal austerity in the Eurozone. \textit{Review of Keynesian Economics, 1}(0), 37-54. \url{https://doi.org/10.4337/roke.2012.01.03}

\bibitem[Zezza(2020)]{zezza2020}
Zezza, G. (2020). Fiscal policies in a monetary union: the eurozone case. \textit{European Journal of Economics and Economic Policies: Intervention, 17}(2), 156-170. \url{https://doi.org/10.4337/ejeep.2020.02.05}

\end{thebibliography}
\end{document}